\documentclass[fleqn,usenatbib]{mnras}
\usepackage{newtxtext,newtxmath}

\usepackage[T1]{fontenc}
\usepackage[flushleft]{threeparttable}
\DeclareUnicodeCharacter{2212}{-}
\DeclareRobustCommand{\VAN}[3]{#2}
\let\VANthebibliography\thebibliography
\def\thebibliography{\DeclareRobustCommand{\VAN}[3]{##3}\VANthebibliography}

\usepackage{lscape}
\usepackage{graphicx}	
\usepackage{amsmath}	
\usepackage{amssymb}	
\usepackage{multicol, blindtext}
\usepackage{float}
\usepackage{times}
\usepackage{lipsum}
\usepackage{caption}
\usepackage{nicematrix}
\usepackage{enumitem}
\usepackage{booktabs}
\usepackage{amsmath}
\usepackage[normalem]{ulem}
\usepackage{lscape}

\usepackage[export]{adjustbox}
\usepackage{placeins}

\usepackage{comment}
\usepackage{subcaption}
\usepackage{xcolor,colortbl}
\usepackage{comment}
\usepackage{empheq} 
\usepackage{float}

\usepackage{xcolor}
\definecolor{lightgreen}{HTML}{90EE90}

\usepackage{soul}

\usepackage{lineno}

\title[Spectral change in BL Lac]{Flaring activity from magnetic reconnection in BL Lacertae
}

\author[S. Agarwal et al.]{S. Agarwal$^1$\thanks{E-mail: \href{mailto:sush.agarwal16@gmail.com}{sush.agarwal16@gmail.com, phd1901221002@iiti.ac.in}},
B. Banerjee$^{2,3}$,  A. Shukla$^1$, J. Roy$^4$, S. Acharya$^{1,5}$,  B. Vaidya$^1$, V. R. Chitnis$^6$, \newauthor S. M. Wagner$^7$, K. Mannheim$^7$, M. Branchesi$^{2,3}$ \\
\textsuperscript{1} Department of Astronomy, Astrophysics and Space Engineering, Indian Institute of Technology Indore, Khandwa Road, Simrol, Indore, 453552 India\\
\textsuperscript{2} Gran Sasso Science Institute, Viale F. Crispi 7, L’Aquila (AQ), I-67100, Italy\\
\textsuperscript{3} INFN - Laboratori Nazionali del Gran Sasso, L'Aquila (AQ), I-67100, Italy\\
\textsuperscript{4} Inter-University Centre for Astronomy and Astrophysics, Pune 411 007, India\\
\textsuperscript{5} Hamburger Sternwarte, Universität Hamburg, Gojenbergsweg 112, 21029 Hamburg, Germany\\
\textsuperscript{6} Tata Institute of Fundamental Research, Homi Bhabha Road, Colaba, 400 005 Mumbai, India\\
\textsuperscript{7} Julius-Maximilians-Universität Würzburg, Fakultät für Physik und 
Astronomie, Institut für Theoretische Physik und Astrophysik, 
Lehrstuhl für Astronomie,\\ Emil-Fischer Str. 31, D-97074 Würzburg, 
Germany}

\date{Accepted XXX. Received YYY; in original form ZZZ}

\pubyear{2022}

\begin{document}
\label{firstpage}
\pagerange{\pageref{firstpage}--\pageref{lastpage}}

\maketitle
\begin{abstract}
The evolution of the spectral energy distribution during flares constrains models of particle acceleration in blazar jets.
The archetypical blazar BL Lac provided a unique opportunity to study spectral variations during an extended strong flaring episode from 2020-2021. 
During its brightest $\gamma$-ray state, the observed flux (0.1-300\,GeV) reached up to $2.15\,\times\,10^{-5}\,\rm{ph\,cm^{-2}\,s^{-1}}$, with sub-hour scale variability. 
The synchrotron hump extended into the X-ray regime showing a minute-scale flare with an associated peak shift of inverse-Compton hump in gamma-rays. 
In shock acceleration models, a high Doppler factor value $>$100 is required to explain the observed rapid variability, change of state, and $\gamma$-ray peak shift.  Assuming particle acceleration in mini-jets produced by magnetic reconnection during flares, on the other hand, alleviates the constraint on required bulk Doppler factor. In such jet-in-jet models, observed spectral shift to higher energies (towards TeV regime) and simultaneous rapid variability arises from the accidental alignment of a magnetic plasmoid with the direction of the line of sight.  We infer a magnetic field of $\sim0.6\,\rm{G}$ in a reconnection region located at the edge of BLR ($\sim0.02\,\rm{pc}$). The scenario is further supported by log-normal flux distribution arising from merging of plasmoids in reconnection region.
\end{abstract}

\begin{keywords}
radiation mechanisms: non-thermal; gamma-rays: galaxies; galaxies:jets; BL Lacertae objects: individual: BL Lac; magnetic reconnection
\end{keywords}

\vspace{-5mm}
\section{Introduction}

The BL Lacertae (BL Lac) is an eponymous blazar at a redshift of 0.069 \citep{1978ApJ...219L..85M} which is usually classified as low peaked BL Lac \citep[LBL][]{2018A&A...620A.185N} with an intermediate BL Lac (IBL) behavior at times \citep{2011ApJ...743..171A}. A peculiar property of the source is the detection of weak $H_{\alpha}$ and $H_{\beta}$ lines underscoring the presence of a feeble broad-line region (BLR) in spite of its classification in the BL Lac class\citep{1996MNRAS.281..737C}. 
Multi-wavelength MWL) studies in flaring and quiescent states require a dominant component of $\gamma$-ray emission from the inverse Compton (IC) upscattering of external seed photons \citep{2011ApJ...730..101A}. Thus, there is a high probability that the BLR serves as a source of seed photons for the electron population in the jets. Emitted high energy (HE) photons are expected to be absorbed and attenuated by the ultraviolet (UV) photons emitted by the BLR and produce a curvature in the HE $\gamma-$ray spectrum \citep{2010ApJ...717L.118P}. Interestingly, the source is a known TeV emitter and has been observed in very-high-energy (VHE; E$>$30\,GeV) $\gamma-$rays by MAGIC, VERITAS \citep{2019A&A...623A.175M, 2013ApJ...762...92A,2018ApJ...856...95A}.
The observed fast TeV variability can be interpreted as a small emission zone close to the black hole magnetosphere \citep{2014Sci...346.1080A}, mini jet-in-jet interaction from magnetic reconnection \citep{2009MNRAS.395L..29G}, star jet interaction \citep{2016MNRAS.463L..26B}, or a two-zone emission region \citep{2011A&A...534A..86T} consisting of a small blob with a large Doppler factor interacting with a larger emission region.
In this work, we strive to elucidate the possible physical processes supporting observed state change in BL Lac during the enhanced activity period. The flow of the paper will be as follows : \S 2 and \S 3 discuss the data reduction and techniques used in analysis. Sections \S 4 and \S 5 cover the results and discussion, respectively.

\vspace{-6mm}
\section{Data Acquisition and Analysis} \label{data_analysis}
The \textbf{\textit{Fermi-}LAT} is a pair-conversion high-energy $\gamma-$ray (HE; 0.1-300\,GeV) telescope on board the Fermi spacecraft. Fermi Science tools and open source {\fontfamily{cmtt}\selectfont Fermipy} package \citep{Wood:2017TJ} are used to analyse source data using the latest instrument response function {\fontfamily{cmtt}\selectfont P8R3\_SOURCE\_V3}.
A circular $15^\circ$ region of interest is considered around the source. \cite{2020NatCo..11.4176S} provides details on the LAT data reduction procedure followed in this work. 

\begin{figure}
    \centering
    \includegraphics[width=1.0\columnwidth]{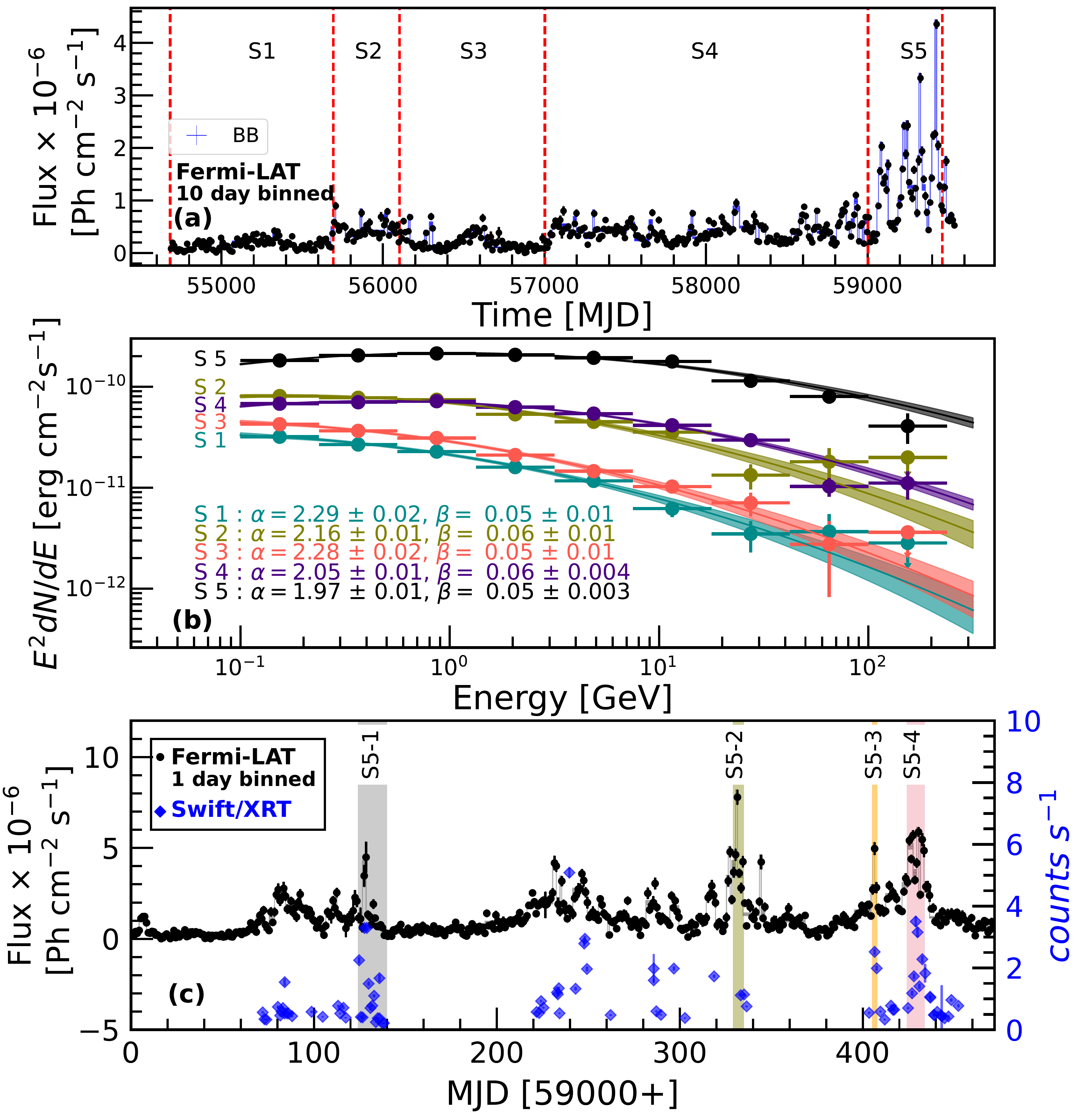}
    \caption{(a) The \textit{Fermi-}LAT LCs of BL Lac for MJD 54683-59473. The red lines categorize 13 years of data into 5 flux states (S1-S5).(b) The high-energy (0.1 - 300\,GeV) spectrum of the 5 states. (c) The 1 day binned \textit{Fermi-}LAT LC of BL Lac for MJD 59000-59478 (S5). The highlighted regions in grey, olive, orange, and pink represent the periods under study (S5-1, S5-2, S5-3, S5-4). Bayesian blocks on top highlight the variable nature of the LC. \textit{Swift-}XRT overall binned LC for MJD 59000- 59478 is plotted alongwith (in blue).}
    \label{fig:fermi_spectrum_different_states}
    \setlength{\belowcaptionskip}{-20pt}
\end{figure}

\begin{figure*}
\centering
\includegraphics[width=0.98\textwidth]{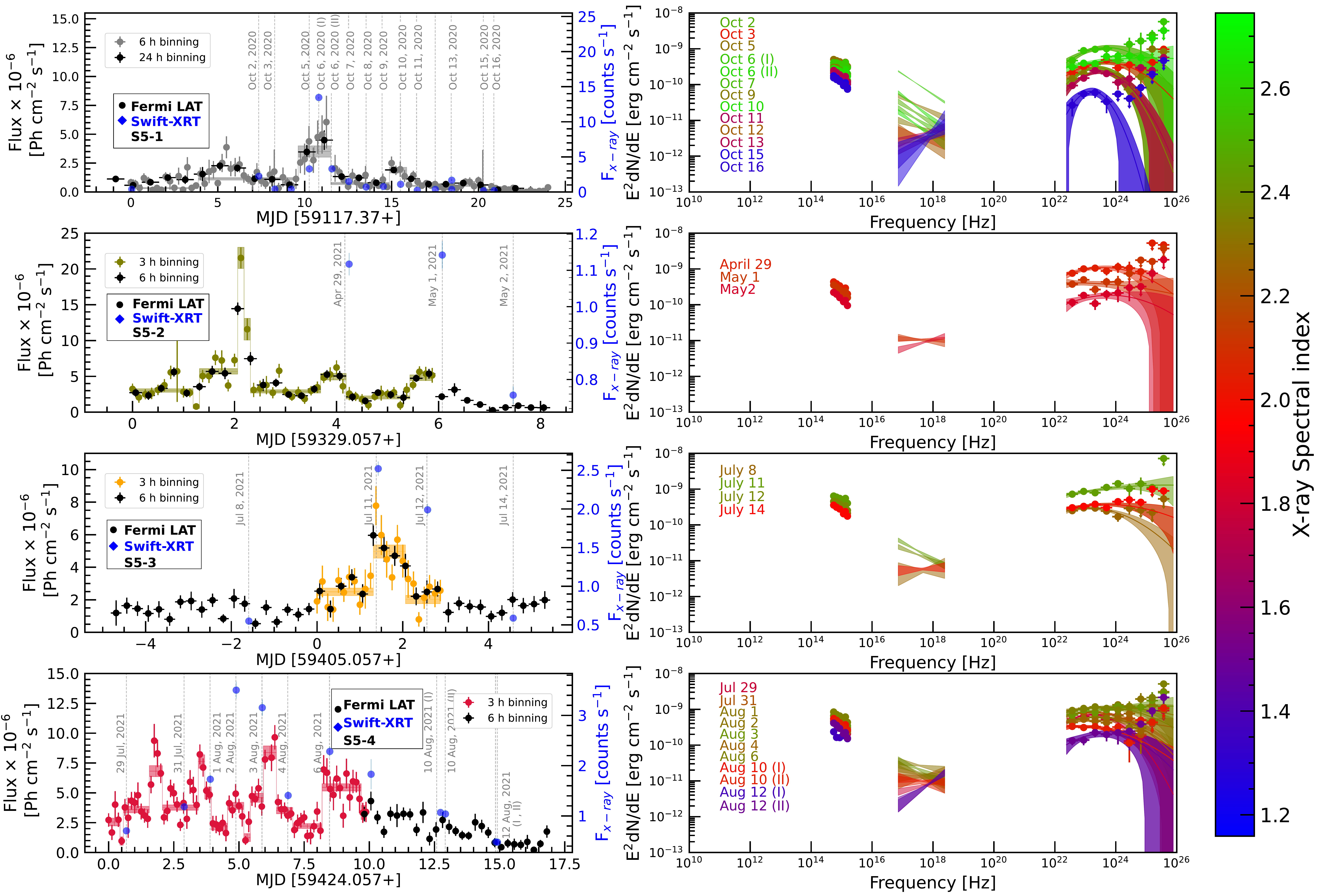}\caption{(Left) The \textit{Fermi-}LAT LC of periods highlighted in Fig. \ref{fig:fermi_spectrum_different_states}c, \ref{fig:fermi_spectrum_different_states}d . The \textit{Swift-}XRT LC is plotted on twin axis in blue. Vertical dashed lines represent the times when the MWL SED is studied (right) for the chosen period.}
 \label{fig:all_flares_sed}
\end{figure*}

We analysed 33 \textbf{\textit{Swift}-XRT} pointed source observations on four time segments corresponding to Fermi flaring episodes (Fig. \ref{fig:fermi_spectrum_different_states}c, \ref{fig:fermi_spectrum_different_states}d). Data reduction to generate light curves, spectrum, ancillary response file, and redistribution matrix file are done using version 1.0.2 of the calibration data base (CALDB) and version 6.29 of the {\fontfamily{cmtt}\selectfont HEASOFT} software for the Photon Counting (PC) and Window Timing (WT) mode data. The data are corrected for the pile-up using the procedures given in website ( \href{https://www.swift.ac.uk/analysis/xrt/pileup.php}{https://www.swift.ac.uk/analysis/xrt/pileup.php} ) 
An annular source region with an outer radius of 30 arc-sec and background region files were extracted farther away from the source using a circular region of radius 50 arc-sec.
The background subtracted spectrum was modelled with an absorbed power law using photoelectric absorption model {\fontfamily{cmtt}\selectfont tbabs} in {\fontfamily{cmtt}\selectfont XSPEC} with a fixed galactic absorption hydrogen column density of $N_H=2.70\times10^{21}\,\rm{cm^{-2}}$ \citep{10.1093/mnras/stab2616} in the source direction. The X-ray flux count rate in Fig. \ref{fig:fermi_spectrum_different_states}c and Fig \ref{fig:all_flares_sed} are from the preliminary analysis of the Swift-XRT data \citep{2013ApJS..207...28S}.

We use simultaneous \textbf{\textit{Swift-}UVOT} observations in all six filters for optical and UV coverage: V ($500$-$600\,\rm{nm}$), B($380$-$500\,\rm{nm}$), U($300$-$400\,\rm{nm}$) and W1 ($220$-$400\,\rm{nm}$ ), M2 ($200$-$280\,\rm{nm}$), W2 ($180$-$260\,\rm{nm}$).
Source counts are extracted from a circle of $5\,\rm{arc-sec}$ radius centered on the source coordinates.
Background counts are derived from a 20 arc-sec radius circular region in a nearby source-free region \citep{10.1093/mnras/stab2616}.
Magnitude and flux are extracted from the generated source and background region files. 
The flux densities for host galaxy in v, b, u, w1, m2 and w2 bands are taken to be $2.89$, $1.30$, $0.36$, $0.026$, $0.020$, $0.017\,\rm{mJy}$  following \cite{2013MNRAS.436.1530R} and subtracted. The host galaxy contaminating the UVOT photometry is $50\%$ of the entire galaxy flux. This contribution is removed from the uncorrected magnitude to obtain the flux free of host galaxy contamination. The galactic extinction is further corrected using E(B-V) value of 0.291 following \cite{2011ApJ...737..103S} with mean galactic extinction laws by \cite{1989ApJ...345..245C}. The corrected magnitude is converted to flux using zero point, and flux density conversion factor from \cite{2008MNRAS.383..627P} and \cite{Roming_2008} respectively.

\vspace{-5mm}
\section{Methods and Techniques}
\textbf{Power Spectral Analysis (PSD):}
We compute the periodogram of $10\,\rm{d}$ binned Fermi-LAT light curve (LC) down to $3\,\rm{h}$ binning to study the temporal variability. Obtained PSD is fitted with a power-law model (PL) of the form PSD$(\nu)\propto\nu^{-k}$, where k and $\nu$ are the spectral index and frequency, respectively. We used the PSRESP method described in \cite{2014MNRAS.445..437M} based on \cite{2002MNRAS.332..231U} to obtain the best fit values of PL parameters. We simulate 1000 LCs having similar flux distribution as the observed \citep{10.1093/mnras/stt764}, accounting for the red noise leakage, and aliasing effects as described in \cite{2022ApJ...927..214G}.\\
\textbf{Bayesian Blocks:}
We use the Bayesian block \citep[BB;][]{2013ApJ...764..167S} algorithm to identify optimal flux states given by constant flux segments of varying duration.
The \textit{point-measurement} fitness function with a \textit{false positive rate} of 5\% is used to find a change point indicating when the flux-state changes to another distinct state \citep{2017ApJ...841..100A}.
The average flux between two change points is considered the flux for the BB. 
The block's flux-uncertainty is the average of each point's flux-uncertainty weighted by the inverse square of flux-uncertainty.
\vspace{-3mm}

\vspace{-3mm}
\section{Results}
The HE (0.1-300\,GeV) LC of BL Lac observed during 13 years has been divided into five states that are marked with vertical dashed lines in Fig. \ref{fig:fermi_spectrum_different_states}a with details provided in Table \ref{tab:psd_results} to study flux and spectral behavior of the source, where a significant increase in flux was seen after MJD 59000. BB analysis showed 94 change points for the 10-day binned LC (Fig. \ref{fig:fermi_spectrum_different_states}a). The PSD analysis on the LC was performed considering different binning from 10 d down to 3 h timescales on 13-year-long \textit{Fermi-}LAT data. The results are tabulated in Table \ref{tab:psd_results}. 

PSDs spectra from 10\,d to 3\,h are consistent with pink noise with index $\sim 1$ of PL function in the 0.1-300\,GeV range. Interestingly, the obtained PSD spectrum is found to be independent of the flux state. The observed consistency in pink noise from 10\,d to 3\,h timescales indicates a similar variability process guiding jet variability regardless of the source's flux state. Furthermore, the variability behavior of individual flaring activity in the flare S5 has been investigated using multiwaveband LCs from \textit{Swift-}XRT and \textit{Fermi-}LAT. 

\begin{figure}
\includegraphics[width=1.0\columnwidth]{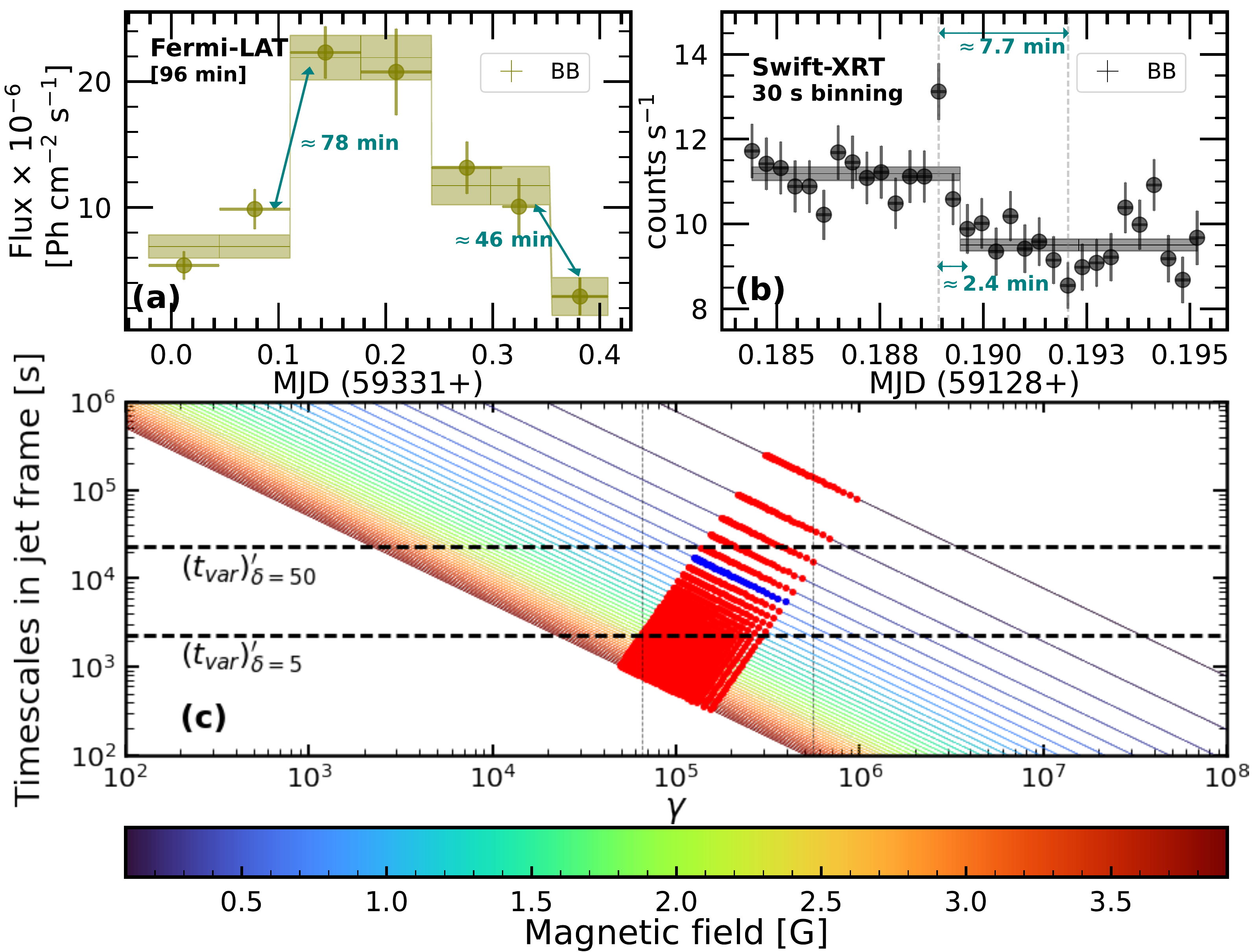}
\caption{(a) Orbit binning LC of BL Lac on April 27, 2021, with BBs with a false positive of 5\% plotted on top. The flux doubling timescales at the point of change of the BB is specified. (b) 30\,s binned \textit{Swift-}XRT LC for BL Lac corresponding to the brightest X-Ray flux observed on Oct 6, 2020. (c) Synchrotron cooling timescale corresponding to different electron energies. The red point represents the timescale for the observed synchrotron emission up to 7.5\,keV. The horizontal black line represents the observed timescales of 7.7\,min in jet frame corresponding to Doppler factor between 5-50.}
\label{fig:figure3}
\end{figure}

We study the spectral evolution for four activity regions in S5: (1) S5-1: MJD\,59120 - 59140 (2) S5-2: MJD\,59329 - 59340 (3) S5-3: MJD\,59400 - 59410 (4) S5-4: MJD\,59420 - 59440 highlighted in Fig. \ref{fig:fermi_spectrum_different_states}c .The choice of the activity regions is based on the availability of dense X-ray data covering the different flux states of corresponding gamma-ray activity. State S5-2 is chosen to account for the variability study during the brightest gamma-ray activity in the source.
The variability timescales are evaluated using,
$t_{var} = (t_2 - t_1) \frac{ln 2}{ln (F_2/F_1)} $,
where $F_2$ and $F_1$ are the fluxes at time $t_2$ and $t_1$ respectively and $t_{var}$ is the flux doubling and halving timescales. 
We checked the shortest variability in X-rays based on the binning of the X-ray LC (10, 15, 20, 25, and 30\,s), where we also checked the combination of the LC points. We observed that a wider binning, such as 30\,s (adopted in this work) results in a higher significance. The resulting significance quoted in this work is a post-trial significance.
The fastest flux variation during S5 was observed on 2020 October, 6 by \textit{Swift-}XRT when the variability of $\Delta t_{var}=7.7 \pm 1.6\,\rm{min}$ was detected with $4.8\sigma$ confidence (post-trial) and a hint of shorter variability of $2.4 \pm 0.9\,\rm{min}$ was also observed with $2.6\sigma$ (post-trial) confidence. These rapid flux changes are also visible as a new BB, see Fig \ref{fig:figure3}b. Simultaneous enhancement in the flux is observed in 0.1-300\,GeV; however, no evidence of coinciding sub-hour variability is found in the LAT data, mostly due to the limited sensitivity of LAT. 
Though, an hour-scale variability with a rise time of 78\,min on 2021 April, 27 and a decay time of 46\,min in the falling part of the flare is observed in orbit binned LC as shown in Fig. \ref{fig:figure3}a for state S5-2. 
\begin{table*}
 \caption{PSD results} 
  \label{tab:psd_results}
  \begin{threeparttable}
    \begin{tabular}{cccccccccc}
    \hline
    Flux state\tnote{1} & Time period\tnote{2} & $T_{obs}$\tnote{3} & $N_{TS>9}/N_{tot}$\tnote{4} & $\Delta T_{min}$\tnote{5} & $\Delta T_{max}$\tnote{6} & $T_{mean}$\tnote{7} & $\alpha \pm \alpha_{err}$\tnote{8} & $p_{\beta}$\tnote{9} & $F_{var} \pm \Delta F_{var}$\tnote{10}\\
     & [day] & [day] &  & [day] & [day] & [day] &  &  & \\

    \hline
    State 1 (S1) & MJD $54683 - 55692$ & $1010$ & $93/101$ & $10$ & $20$ & $10.89$ & $0.81 \pm 0.37$ & $0.80$ & $0.52 \pm 0.03$\\
    State 2 (S2) & MJD $55693 - 56103$ & $410$ & $41/41$ & $10$ & $10$ & $10$ & $0.43 \pm 0.66$ & $0.98$ & $0.30 \pm 0.02$\\
    State 3 (S3) & MJD $56103 - 57003$ & $900$ & $83/90$ & $10$ & $30$ & $10.86$ & $1.07 \pm 0.52 $ & $0.65$ & $0.71 \pm 0.03 $ \\
    
    S1+S2+S3 & MJD $54683 - 57003$ & $2320$ & $216/231$ & $10$ & $30$ & $10.75$ & $1.17 \pm  0.34$ & $0.51$ &  $ 0.69 \pm 0.01 $\\

    State 4 (S4) & MJD $57003 - 59003$ & $2000$ & $195/195$ & $10$ & $50$ & $10.26$ & $0.94 \pm 0.23$ & $0.36$ & $0.42 \pm 0.01 $\\
    
    State 5 (S5) & MJD $59003 -59463$ & $460$ & $46/46$ & $10$ & $10$ & $10$ & $1.24 \pm 0.62$ & $0.22$ & $0.64 \pm  0.01 $\\
    State 5 (S5) & MJD $59000 -59463$ & $464$  & $447/464$ & $1$ & $4$ & $1.04$ & $1.21 \pm 0.21$ & $0.46$ & $ 0.84 \pm 0.01 $\\
    S5-4 & MJD $59420 - 59440$ & $10$ & $79/79$ & $0.125$ & $0.25$ & $0.13$ & $0.77 \pm  0.34$ & $0.14$ & $0.41 \pm 0.02$\\
    S1+S2+S3+S4+S5 & MJD $54683 -59463$ & $4770$ & $457/472$ & $10$ & $50$ & $10.44$ & $1.24 \pm 0.29$ & $0.84$ & $1.04 \pm  0.01$\\
    \hline
    \end{tabular}
    \end{threeparttable}
    \begin{tablenotes}
    \small
 \item Note: (1)  The flux states based on flux levels (S1, S2 and S3 are combined to improve the statistics as a number of points are less in S2 and S3) (2) time period (3) Total exposure (4) Fraction of points having TS greater than 9 (5) Minimum sampling interval in observed LC (6) Maximum sampling interval  in observed LC (7) Mean Sampling time interval i.e. total observation time over a number of data points in that interval (8) The power law index for the power law model of PSD analysis (9) P value corresponding to the power law model. The power law model is considered a bad fit if $p_{\beta}\leq0.1$ as the rejection confidence for such model is $>90\%$ (10) Fractional variability
\end{tablenotes}
\end{table*}

The flux profiles for different states of BL Lac (S1-S5) are estimated using \citealt{2021MNRAS.504.1427A}. 
The best-fit model is selected based on a non-overlapping distribution (outside 1$\sigma$) of the Akaike Information Criterion (AIC) derived from 1000 simulated LCs from a Gaussian, where the mean and width are the observed flux and flux uncertainty, respectively.
For S3 and S5, Log-Normal flux distribution is preferred over a Gaussian. The source shows larger variability for S3 and S5 with the values of fractional variability \citep[F$_{var}$;][]{2003MNRAS.345.1271V} as 0.70$\pm$0.03 and 0.64$\pm$0.01, respectively. A Gaussian distribution is preferred in S1, where no prominent flare has been observed. The log-normality of flux is also reported in several other blazars in X-rays to VHE $\gamma$-rays (\citealt{2021MNRAS.504.1427A} and ref. therein) as a consequence of multiplicative processes responsible for the variability. However, \cite{2020ApJ...895...90S} shows in the general case that multiplicativity is not necessarily needed to obtain an rms-flux correlation and that such correlations have been obtained in specific examples through purely additive processes \citep{2012A&A...548A.123B}.

To study the spectral evolution over 13\,years, the HE LAT spectra of different states  are fitted with a log parabola model, parametrized as $\frac{dN}{dE}=N_\circ \left( \frac{E}{E_b} \right)^{-(\alpha + \beta(log(E/E_b)))}$
where $E_b$ was fixed to 4FGL catalog value of $0.7$\,GeV. The best fit spectral parameters are listed in Fig. \ref{fig:fermi_spectrum_different_states}b. The observed spectral indices ($\alpha$) show a trend with the increasing flux for the five states; however, the curvature parameter ($\beta$) is found to be consistent for each state. In addition, the peak of the HE spectrum shifts to higher energy and is observed at 1\,GeV for S5, which is the period with the brightest $\gamma$-ray emission.
MWL spectral evolution consisting of UV, X-rays and GeV data is studied during the four activity periods.
The epochs with simultaneous observation in \textit{Swift-}XRT and \textit{Fermi-}LAT are chosen based on the BBs covering the Swift observation epoch and are shown by vertical dashed lines in Fig. \ref{fig:all_flares_sed}. Simultaneous MWL SEDs are
shown on the right side of Fig. \ref{fig:all_flares_sed}. The X-ray spectrum is fitted using the absorbed power-law model described in \S~\ref{data_analysis}. 
{The source exhibits a softer when brighter behavior in the energy range of 0.2-10\,keV in S5.} \\
An evident state change in the X-ray regime is flagged by X-ray emission observed in the second hump of the spectrum during the low flux state, which evolves into softer X-ray emission via the synchrotron process during the enhanced flux states. This shift is accompanied by hint of 
 shift of the \textit{Fermi}-LAT spectrum to higher energies, as indicated by an apparent shift of the second hump's peak to higher energies and the simultaneous detection of the highest energy photons (HEP) for the studied state.
 This effect is especially noticeable in the flare of S5-1, S5-4. For S5-1, the X-ray emission lies in the rising part of the EC hump from 2020 October 11 to 2020 October 16, in contrast to the observed X-ray emission via synchrotron process from 2020 October 2 to 2020 October 10. The stacked LAT data based on the hardness ratio in the X-rays, softer and harder than $-2$ results in the peak of LAT spectra at energies $1.06 \pm 0.21\,\rm{GeV}$ to $1.97 \pm 0.05\,\rm{GeV}$, indicating the HE peak shifts to higher energies as the X-ray spectra becomes harder. 
 For the observed shifted HE peak, the detected HEP ranged from $7.6\,\rm{GeV}$ to $53.6\,\rm{GeV}$ respectively.
Similarly, for period S5-4, we observe a transition of X-ray emission via synchrotron process from the EC process as the flux evolves from 2021 July, 29 to 2021 August, 2. As the flux decays further on 2021 August, 12 the spectrum shifts to the second hump.
This is visible in the stacked LAT spectrum with a peak at $0.67 \pm 0.14\,\rm{GeV}$ during the periods of hard X-ray spectrum to a shift in peak at $1.84 \pm 0.63\,\rm{GeV}$ during periods of softer X-ray spectrum. The shift is further supported by the detection of HEP of energies $71\,\rm{GeV}$ to $114\,\rm{GeV}$, respectively during the stacked periods. A hint of a similar shift in S5-3 is highlighted by the detection of HEP of $172\,\rm{GeV}$ during July 11, 2021 where the X-ray spectrum lies in the first hump in contrast to the HEP of $50\,\rm{GeV}$ during July 14, 2021 where the X-ray spectrum is significantly harder.
The spectral shift is shown in Fig. \ref{fig:all_flares_sed}. 
\vspace{-7mm}

\section{Discussion}
BL Lac's flux levels are found to be variable and evolved with time, and its HE spectrum (0.1-300 GeV) can be explained by the log-parabola model. The \textit{Fermi-}LAT spectral index ($\alpha$) gets harder with increasing flux, suggesting fresh or re-accelerated electrons. The spectrum's curvature parameter ($\beta$) does not change over 13\,years despite a significant flux change, suggesting a similar influence of external UV photons within or at the edge of the BLR \citep{2010ApJ...717L.118P} on emitted photons in the jet. The observed Lyman $H_{\alpha}$ lines suggest a weak BLR since from the standard scaling relation, Luminosity $L_{BLR}= 2.5 \times 10^{42}$ $\rm{erg/s}$ and $R_{BLR}=2\times 10^{16}$ $\rm{cm}$ \citep{2009MNRAS.397..985G}.

The source is known to be variable in multiple waveband \citep{2020ApJ...900..137W} . It was found in high activity in 2020-2021 and multiple episodes of state change, where X-ray emission shifts from the second to first SED hump. For the first time, minute-scale X-ray variability was found simultaneous with a rare shift of the X-ray emission to the first hump. Moreover, rapid variability and X-ray state change were accompanied by a simultaneous shift of IC peak to the higher energies in activity region S5-1 and S5-4. Such events are extremely rare in blazars and help constrain emission and particle acceleration models. For the brightest $\gamma$-ray flux observed on MJD\,59331, the orbit binning reveals a sub-hour variability of $46\,\rm{min}$, consistent with observed TeV variability \citep{2013ApJ...762...92A}. In addition, the correlation of flux-rms vs. flux and the dominance of log-normal flux distribution could indicate a multiplicative effect associated with the accretion process \citep{2005MNRAS.359..345U}. 
A minijet-in-jet model can also be a possible explanation for these observations \citep{2012A&A...548A.123B}. Similar PSD for categorised states suggests a similar variability process in the Fermi band. The quasi-simultaneous detections of TeV emission, rapid variability, peak-shift and X-ray observation at the first hump, and Log-Normal distribution poses substantial challenges to the shock-in-jet model \citep{10.1046/j.1365-8711.2001.04557.x}.

Corresponding to MJD 59128, during the period of brightest X-ray flux, a minimum variability time of $7.7 \pm 1.6\,\rm{min}$ with a post-trial significance of 4.8$\sigma$ in X-ray LCs is detected. This corresponds to an emission region located within the BLR at $ 2.9 \times 10^{15} \rm{cm}$ if the emission region covers the entire cross-section of the jet. We also found hints of a shorter variability timescale of  $2.4 \pm 0.9\,\rm{min}$ ($2.6\sigma$, 
 post-trial) in 30\,s binned \textit{Swift-}XRT data. Similar results are reported by \cite{10.1093/mnras/stab2616,2021arXiv210812232S}. A sub-hour variability of $46\pm24\,\rm{min}$ is observed on MJD\,59331 during the brightest $\gamma$-ray state of the source. \cite{2022arXiv221000799P} hinted at a minute scale GeV $\gamma-$ray variability during this giant $\gamma-$ray outburst.

The extension of the synchrotron spectrum up to $\sim$7.5\,keV during the high-flux states and hardening of the X-ray spectra during the low flux state hints at a selective viewing angle during the flare ($v_{syn} \propto \gamma^2 B \delta$) or a significant particle acceleration process. The observed 7.5\,keV photons provides a signature of the maximum energy of the accelerated electrons. Using synchrotron cooling timescales, $\tau = \frac{3 m_{e} c}{4 \sigma_T \gamma U}$, from Eq. 12 in \citealt{10.1111/j.1365-2966.2008.14270.x} and frequency of emitted synchrotron photons, $\nu_s = 4.2 \times 10^6 \gamma^2 B^{\prime} \frac{\delta}{1+z}$\,Hz \citep{2021arXiv210200919C}, we constrain $\gamma^2 B^{\prime} \delta = 4.3 \times 10^{11}$\,Hz, where U=U$_{\rm mag}$=B$^{\prime 2}/8\pi$ for synchrotron losses. The observed timescale of 7.7\,min is translated into the jet frame by using a Doppler factor between 5-50. The electron energies responsible for the observed emission of 7.5\,keV are found to be $\gamma= 6.5\times10^4-5.5\times10^5$. This limits the magnetic field to be within 0.3 - 2.2\,G (see Fig.\ref{fig:figure3}c). 

The shock-in-jet scenario and recollimation shock demands a Doppler factor > 100 for the observed luminosity from an emission region corresponding to
observed minute scale variability \citep{2009ApJ...699.1274B}.
Such high Doppler factor values contradict the values in kinematic studies of parsec scale jets and also from magnetohydrodynamical (MHD) models of the jets \citep{2005AJ....130.1418J}. 

A possible origin for extended X-ray emission up to 7.5\,keV along with observed fast variability and the apparent shift into the second hump could be associated with the preferred alignment of the emission along the line of sight \citep{2021ApJ...912...40M} through jet-in-jet scenario. Substantial dissipation takes place when reconnection timescales become equal to expansion timescales of jet at distances $R_{diss} \simeq \Gamma^2 r_g / \varepsilon$ here $\varepsilon$ parametrizes the reconnection rate \citep{10.1093/mnras/stt167}.
Thus the dissipation takes place at $R_{diss} = 4.74 \times 10^{16}\,\rm{cm} = 1012\,\rm{r_g}$ from the central engine, close to the outer boundary of BLR. At the sight of reconnection, magnetic energy is transferred to the particles and results in plasmoid formation \citep{2019MNRAS.486.1548M}.
We expect  an enhanced emission and a shift in SED to the higher energies due to Doppler enhancement caused by selective orientation of plasmoid in observer's line of sight.
However, when the source fades back to the low state, post flare, or when the plasmoid is no longer in line of sight, the Doppler boost fades away, and the SED shifts to lower energies. 

Considering the jet aligned in line of sight, $\Gamma_{j}$=10, we compute the Doppler factor of a large plasmoid to be $\delta_{p}$=40. The emission from the entire reconnection region results in envelope emission, which is significantly lower than the emission from the mini-jets specifically aligned in the observer's direction. The characteristic size $l'$ is estimated from the envelope timescale $t_{env}$ as $l'=t_{env} \Gamma_j \varepsilon c \sim 5.1 \times 10^{15}\,\rm{cm}$. The plasmoid responsible for the minute scale flare grows up to $10\%$ ($f=0.1$) of the reconnection region. The rise/decay time of the minute scale flare on top of the envelope emission is given by $t_{flare}=fl'/\delta_{p} c \sim 425\,\rm{s}$. Total envelope and plasmoid luminosities which are responsible for envelope and fast-flare emission respectively in jet-in the-jet model can be expressed as $ L_{env} =2\Gamma^{2}_{j}\delta^{2}_{p}l^{'2} U^{'}_{j}\varepsilon c$ erg/s \& $L_{P} =4\pi f^{2}l^{'2} U^{''}_{p} c\delta^{4}_{p}~$\,erg/s respectively. Here $\varepsilon$ is the reconnection rate, U$^{'}_{j}$ is the energy density at the dissipation zone in the co-moving frame of the jet, U$^{''}_{p}$ is the energy density of the plasmoid in its co-moving frame. We use $U^{'}_{j} = U^{''}_{p}/2$  as in \cite{10.1093/mnras/stt167}.The isotropic envelope and plasmoid luminosity is found to be $L_{env}=3.6\times10^{44}\,\rm{erg/s}$ and $L_{p}=7.2\times10^{45}\,\rm{erg/s}$ (for $B^{\prime}=0.6\,\rm{G}$) respectively. For such a magnetic field, electron energies for the observed cooling timescales are within $1.2 \times 10^{5}-4 \times 10^{5}$.

We conclude that the SED variations and their time scales reported are in line with a scenario which involves a flaring and a steady component. Magnetic reconnection gives rise to impulsive particle acceleration in mini-jets associated with a stochastic flaring component. Growing modes of kink instability can lead to magnetic  reconnection beyond the edge of the BLR out to the parsec scale. Such kink instabilities have been located by \citealt{Jorstad2022} beyond  5\,pc fueling observed optical activity and co-spatial $\gamma$-ray  emission through Synchrotron Self Compton. Close to the edge of the BLR, the SED is dominated by inverse-Compton scattering off external optical BLR photons \citep[e.g.][]{2019A&A...623A.175M}.

\vspace{-6mm}
\section*{Acknowledgements}
SA acknowledges Arti Goyal for useful discussion on manuscript. We thank the referee, Jonathan Biteau, for constructive feedback during the review process. BB and MB acknowledge financial support from MUR (PRIN 2017 grant 20179ZF5KS). 
\vspace{-6mm}
\section*{Data Availability}
The data used in this article will be shared upon reasonable request.


\vspace{-6mm}
\bibliographystyle{mnras}
\bibliography{main}

\appendix

\label{lastpage}
\end{document}